\documentclass[twocolumn,showpacs,preprintnumbers,amsmath,amssymb]{revtex4-1}
\usepackage{epsfig}
\usepackage{graphicx}
\usepackage{epsfig}
\usepackage{rotating}
\usepackage{amssymb}
\usepackage{amsmath}
\usepackage{graphicx}

\def\beq{\begin{eqnarray}}
\def\eeq{\end{eqnarray}}

\newcommand{\be}{\begin{equation}}
\newcommand{\ee}{\end{equation}}
\newcommand{\bea}{\begin{eqnarray}}
\newcommand{\eea}{\end{eqnarray}}

\begin{document}

\title{Inverse Ising inference using all the data}

\author{Erik Aurell}
 \altaffiliation[Also at ]{Aalto University School of Science, Helsinki, Finland
}
 \email{eaurell@kth.se}
\affiliation{ACCESS Linnaeus Centre, KTH, Stockholm, Sweden\\
and Dept.~Computational Biology, AlbaNova University Centre, 106 91 Stockholm, Sweden}
\author{Magnus Ekeberg}
\email{ekeb@kth.se}
\affiliation{Engineering Physics Program, KTH Royal Institute of Technology, 100 77 Stockholm, Sweden}

\begin{abstract}
We show that a method based on logistic regression, using all the data, solves the inverse Ising problem far better than mean-field calculations relying only on sample pairwise correlation functions, while still computationally feasible for hundreds of nodes. The largest improvement in reconstruction occurs for strong interactions. Using two examples, a diluted Sherrington-Kirkpatrick model and a two-dimensional lattice, we also show that interaction topologies can be recovered from few samples with good accuracy and that the use of $l_1$ regularization is beneficial in this process, pushing inference abilities further into low-temperature regimes.
\end{abstract}

\keywords{Inverse Ising problem, pseudolikelihood, logistic regression}
\maketitle


{\it Introduction:}
When analyzing systems of interacting elements from data, disentangling direct 
from indirect interactions is an intrinsically complex task.
Versions of this problem come about naturally in biology, sociology, neuroscience and many other fields, and are
bound to become more and more important as the amount and diversity of data on large systems continue to grow. 
In the Ising model, which has served as a basic starting point for studying such situations in 
applications~\cite{Bialek2006,Lezon2006,Cocco2009}, a set of binary variables 
$\boldsymbol{\sigma}=\{\sigma_1, ... ,\sigma_N\}$, $\sigma_i=\pm 1$, has the distribution
\begin{equation}
\label{ising_dist}
P(\boldsymbol{\sigma})= \frac{1}{Z} \exp\left(\beta \sum\limits_{i} {h_i \sigma_i}+\beta \sum\limits_{i<j}{J_{ij}\sigma_i\sigma_j}\right)
\end{equation}
where $Z$ is the partition function, $\beta=1/T$ the inverse temperature, $h_i$ are the external fields and $J_{ij}$ the pairwise couplings (representing direct interactions). 
Given magnetizations $m_i=\langle \sigma_i \rangle$ and pairwise correlations $c_{ij}=\langle \sigma_i\sigma_j \rangle-m_im_j$
the probability distribution which maximizes the entropy has the Ising model form. The standard \textit{inverse Ising problem} means to 
compute (approximately, efficiently, or according to other criteria) the parameters $h_i$ and $J_{ij}$ from
observed $m_i$ and $c_{ij}$. 
The practical interest in inverse Ising, in the context of the present and future data-rich world, is to use it
as an information extraction tool superior to measuring correlations. For example, Ising models can explain the higher order correlations observed in networks of neurons \cite{Tkacik2009} and, extending the number of states from two to 21, spectacular success has
been achieved in predicting protein structure by inferring directly interacting residues (amino acids) \cite{SchugPNAS2009,WeigtPNAS2009,Morcos-2011}.
In this Letter, we address the following two questions: \textit{(i)} can one do better by keeping all the data for reconstruction and not only empirical pairwise correlation functions, and \textit{(ii)} can such a method be implemented
in a computationally efficient manner? The answer is positive on both accounts, using a method
inspired by the regularized logistic regression of Wainwright, Ravikumar, and Lafferty~\cite{RavikumarWainwrightLafferty10}.
We show, in particular, that keeping all the data greatly improves reconstruction of an Ising model in 
the important parameter region of strong interactions.

{\it Maximum likelihood and computability:} 
Given $B$ independent observations $\{\boldsymbol{\sigma}^{(k)}\}_{k=1}^B$ all drawn from
(\ref{ising_dist}), the log-likelihood function is 
\begin{eqnarray}
\label{log-likelihood}
l(\{h_i\},\{J_{ij}\};\{\boldsymbol{\sigma}^{(k)}\}_{k=1}^B) &=& \beta\sum_i h_i m^{(B)}_i +\nonumber\\
                                                        \beta \sum_{i<j} J_{ij}(m^{(B)}_im^{(B)}_j+c^{(B)}_{ij}) &-& \log Z
\end{eqnarray} 
where $m^{(B)}_i$ and $c^{(B)}_{ij}$ are the empirical first and second moments.
In (\ref{ising_dist}), the averages of the functions multiplying the model parameters
are sufficient statistics~\cite{darmois35,pitman36,koopman36}, which in the case at hand means that inference of the biases
$h_i$ and the interaction strengths $J_{ij}$ cannot be done better using 
all the $B$ samples ($NB$ data points), 
than by observing just $m^{(B)}_i$ and $c^{(B)}_{ij}$ ($\frac{N(N+1)}{2}$ data points). 
The optimal estimates (in a maximum likelihood sense) are given by $\partial_{h_i}\log Z=\beta m^{(B)}_i$ and $\partial_{J_{ij}}\log Z=\beta [m^{(B)}_im^{(B)}_j+c^{(B)}_{ij}]$. 
Boltzmann learning \cite{Ackley85alearning} uses Monte Carlo (MC) sampling to compute these gradients of the partition function with
, in principle, unbounded accuracy, but is computationally tractable only for very small systems (although faster versions of this procedure have been introduced, see, \textit{e.g.}, \cite{Broderick} or \cite{CarreiraHinton}).
A whole series
of approximations, reviewed in~\cite{Frontiers}, have therefore been developed by expanding in high temperature (small interactions), large external fields, or other
parameters \textit{cf.} (naive) mean field (nMF)~\cite{Rodriguez97efficientlearning}, TAP~\cite{Rodriguez97efficientlearning}, small-correlation expansion (SCE) \cite{SessakMonasson}
and have been further extended using the fluctuation-dissipation theorem~\cite{MezardMora, Marinari,AurellOllionRoudi10}.
It is well-established that all these approximate methods are not accurate when the number of samples
is small, nor when the interactions are strong (temperature is low). However, a recent method based on expansion of the system into "clusters" (the contributions of which to the estimates of $\{\boldsymbol{\mathrm{h}},\boldsymbol{\mathrm{J}}\}$ are included or discarded depending on their entropy share) manages to select correctly the parameters from few samples in various low-temperature settings \cite{CoccoMonasson2011}, questioning these limitations. Another promising candidate called minimum probability flow, recently introduced in \cite{PhysRevLett.107.220601}, has been shown to very efficiently recover Ising parameters for a two-dimensional grid. Its performance on more general systems, in particular strongly correlated ones, is an interesting and open question.


{\it Pseudolikelihood maximization (without regularization):} 
The conditional probability of one variable $\sigma_r$ given all the others $\boldsymbol{\sigma}_{\backslash r} =(\sigma_1, ... ,\sigma_{r-1},\sigma_{r+1}, ... ,\sigma_N)$ is
\begin{equation}
\label{cond_prob}
P_{\{\boldsymbol{\mathrm{h}},\boldsymbol{\mathrm{J}}\}}(\sigma_r|\boldsymbol{\sigma}_{\backslash r})= \frac{1}{1+e^{-2\beta \sigma_r [h_r+\sum_{i\neq r}{J_{ir}\sigma_i}]}} \
\end{equation}
where we take $J_{ir}$ to mean $J_{ri}$ when $i>r$. If $\sigma_r$ by itself is considered a dependent variable, and the complementary set $\boldsymbol{\sigma}_{\backslash r}$ is taken as independent variables, 
then the maximum likelihood estimates of the parameters $h_r$ and $\boldsymbol{\mathrm{J}}_{r}=\{J_{ir}\}_{i \neq r}$,  given $B$ samples, minimize 
\begin{equation}
\label{f_def}
f_r(h'_r,\boldsymbol{\mathrm{J}}'_r)=-\frac{1}{B} \sum_{k=1}^B {\ln P_{\{\boldsymbol{\mathrm{h}}',\boldsymbol{\mathrm{J}}'\}}(\sigma_r^{(k)}|\boldsymbol{\sigma}_{\backslash r}^{(k)})} \ .
\end{equation} 
Minimizing these functions $f_r$ for all $r$ simultaneously, which we call pseudolikelihood maximization, is not the same as maximizing the total log-likelihood~(\ref{log-likelihood}). For
example, it typically gives different estimates $J^{*,i}_{ij}$ and $J^{*,j}_{ij}$ depending on if $\sigma_i$ or $\sigma_j$ is considered
the dependent variable. We will for definiteness sake always take
$J^*_{ij}=\frac{1}{2}\left(J^{*,i}_{ij}+J^{*,j}_{ij}\right)$.
Alternatively, one could minimize the sum of all $f_r$ while requiring $J^{*,i}_{ij}=J^{*,j}_{ij}$.
When the number of samples is large we can substitute sample average with ensemble average, and write
\begin{eqnarray}
\label{f_Binf}
f_r(h'_r,\boldsymbol{\mathrm{J}}'_r)&\approx& \left\langle \ {-\ln \left(P_{\{\boldsymbol{\mathrm{h}}',\boldsymbol{\mathrm{J}}'\}}(\sigma_r|\boldsymbol{\sigma}_{\backslash r})\right)} \ \right\rangle \nonumber\\
                          &=&\sum_{\boldsymbol{\sigma}} {\ln \left(1+e^{-2\beta \sigma_r [h'_r+\sum_{i\neq r}{J'_{ir}\sigma_i}]}\right) P_{\{\boldsymbol{\mathrm{h}},\boldsymbol{\mathrm{J}}\}}(\boldsymbol{\sigma})}\ , \nonumber\\
\end{eqnarray} 
with equality expected in the limit. Necessary maximum likelihood conditions (for one of the conditional probabilities) are then
\begin{equation}
\label{d}
\frac{\partial f_r}{\partial {J'_{sr}}}(h'_r,\boldsymbol{\mathrm{J}}'_r)=\sum_{\boldsymbol{\sigma}} \frac{-2 \beta \sigma_s \sigma_r}{e^{2\beta \sigma_r [h'_r+\sum_{i\neq r}{J'_{ir}\sigma_i}]}+1} P_{\{\boldsymbol{\mathrm{h}},\boldsymbol{\mathrm{J}}\}}(\boldsymbol{\sigma}) =0
\end{equation}
and similarly for the variation with respect to an external field. At the true parameters these equations hold, since
\begin{eqnarray}
\label{cons}
\frac{\partial f_r}{\partial {J'_{sr}}}(h_r,\boldsymbol{\mathrm{J}}_r)=& \nonumber \\
\frac{-\beta}{Z\{\boldsymbol{\mathrm{h}},\boldsymbol{\mathrm{J}}\}} \sum_{\boldsymbol{\sigma}} &{\sigma_s \sigma_r \frac{e^{\beta \sum_{i \neq r}{h_i \sigma_i}+\beta \sum_{\underset{i,j \neq r}{i<j}}{J_{ij}\sigma_i\sigma_j}}}{cosh(\beta \sigma_r [h_r+\sum_{i\neq r}{J_{ir}\sigma_i}])}}=0 \ ,  \ 
\end{eqnarray}
where the expressions vanish because each state for which $\sigma_r=1$ has exactly one opposing state for which $\sigma_r=-1$, contributing equally in size. Assuming this stationary point is a minimum we can locate, the pseudolikelihood approach to inferring an Ising model is exact in the limit of large sample size, and is in this sense qualitatively different
from other approximate inverse Ising schemes.

{\it Pseudolikelihood maximization with $l_1$ regularization:}
Ravikumar, Wainwright and Lafferty in~\cite{RavikumarWainwrightLafferty10} introduced a $l_1$-regularized version of the pseudolikelihood
approach, \textit{i.e.} one where the functions to be minimized are
$\left[f_r(h'_r,\boldsymbol{\mathrm{J}}'_r) + \lambda ||\boldsymbol{\mathrm{J}}'_r||_1\right]$ with some
penalty parameter $\lambda>0$.
$l_1$(absolute value) regularization is widely used to recover sparse signals~\cite{Tibshirani96,DonohoElad03,NYAS:NYAS03764}, in situations
where a large fraction of parameters is known to be zero, but not which parameters. The numerical minimization can be done efficiently
using convex programming, such as the interior point method of Koh, Kim and Boyd~\cite{KohKimBoyd07}, which we have used below.

 \begin{figure}[t]
\centering
\includegraphics{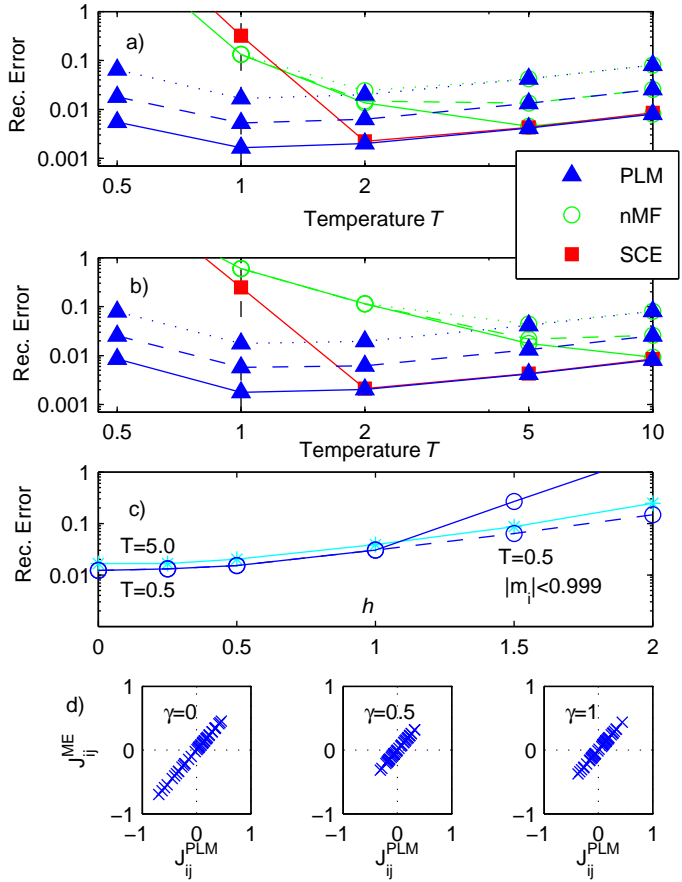}
\caption{(Color online) a) and b) show reconstruction errors of pseudolikelihood maximization (PLM), nMF and SCE
versus temperature for a) fully ($p=1$) and b) sparsely ($p=0.1$) connected SK-systems of size $N=64$. The number of MC samples used are $10^6$ (dotted), $10^7$ (dashed) and $10^8$ (continuous). c) shows reconstruction errors of PLM as functions of external field strength for a SK system of size $N=10$ using $10^6$ samples for two different temperatures. The dashed curve is obtained for $T=0.5$ by excluding parameter sets where one or more empirical $|m_i|>0.999$. d) compares parameter estimates between Boltzmann learning ($J_{ij}^{ME}$) and PLM ($J_{ij}^{PLM}$) using data generated from a distribution with Hamiltonian $-(1-\gamma)\sum_{i<j}{J_{ij}\sigma_i\sigma_j}-\gamma\sum_{i<j<k}{J_{ijk}\sigma_i\sigma_j\sigma_k}$. The system parameters used are $N=10$, $B=10^6$, $T=2$, and all interaction parameters are drawn from a $N(0,\frac{1}{\sqrt{N}})$-distribution.}
\label{fig:highSampleSimulations}
\end{figure} 

{\it Results for high-quality data:}
We minimized (\ref{f_def}) using Newton decent for several values of $B$ in the setting of the dilute Sherrington-Kirkpatrick (SK) model~\cite{SherringtonKirkpatrick75}, a commonly used test bench for comparing performances of inverse Ising solvers. 
Every $J_{ij}$ is thus nonzero with probability $p$, and if so drawn from a Gaussian distribution with zero mean and variance $1/c$, $c=pN$. 
External fields were first taken as zero. 
Reconstruction error was measured by
$\Delta=\frac{1}{1/\sqrt{N}} \left\langle (J^*_{ij}-J_{ij})^2 \right\rangle^{1/2}$. 
For comparison we also applied nMF, TAP and two versions of the small-correlation expansion: the general result of \cite{SessakMonasson}, as well as their higher order zero-magnetization version. The latter is currently the best performing approximate method tested on the zero-field SK model. Figure~\ref{fig:highSampleSimulations}a)-b) shows simulation results for $N=64$ compared to nMF, \textit{i.e.} $J_{ij}^{nMF}=-\frac{1}{\beta}\left(\boldsymbol{c}^{-1}\right)_{ij}$. We also include one curve (for $10^8$ samples) for the best performing competing method, which indeed turned out to be the higher order SCE. The curves are the averages of five different parameter sets, yielding error bars small enough to be omitted. MC sampling was performed using a warmup time of $10^7 \cdot N$ spin updates and a sampling frequency of one observation every $10 \cdot N$ updates. Evidently, pseudolikelihood maximization outperforms nMF and SCE in the low-temperature region. As $T$ approaches one from above (toward the spin glass phase), nMF and SCE start to perform poorly, while our logistic regression algorithm appears unaffected. Lowering the temperature further to $T=0.5$, where indeed all approximate methods tested on this example to date are unusable, pseudolikelihood maximization continues to function. This makes it the first tested method to reconstruct successfully in this strongly correlated region of the SK model. As the temperature increases, performance is limited by the finiteness of $B$ rather than by the method choice. In this limit (right part of the curves in Figure~\ref{fig:highSampleSimulations}a)-b)) reconstruction error $\Delta$ follows $\sim \frac{1}{\sqrt{B}}$ for all methods, but for pseudolikelihood maximization this seems to hold for all temperatures (parallel curves in Figure~\ref{fig:highSampleSimulations}a)-b)) . The switch to sparse $\boldsymbol{\mathrm{J}}$ clearly worsens the performance of nMF,
but does not seem to affect the pseudolikelihood scheme or SCE much if at all. The results for system sizes $N=16$ and $N=128$ were similar (data not shown). 

In applications, external fields typically are not zero, and there will usually also be some interdependence between samples. It is thus natural to ask whether the good results for $T<1$ are maintained when relaxing these assumptions. To assess the robustness of pseudolikelihood maximization against biases, we let all $\beta h_i = h$ and observed the reconstruction error as $h$ was increased. Figure~\ref{fig:highSampleSimulations}c) shows results for a SK model with N=10 using $10^6$ samples in the cases $T=5$ and $T=0.5$. 
For both temperatures, a moderate change in the fields has no effect on performance. 
For weak interactions ($T=5$), strong magnetizations yield only a modest increase in error (for the $h=2$ case all means are $m_i \approx 0.96$). When both couplings and fields are strong, the situation is more delicate. 
An examination of the data showed that the decrease in performance for $T=0.5$ and $h>1$ stemmed from a few inference runs having huge errors. For instance, at $h=1.5$, over $90\%$ of the runs still showed errors $\Delta <0.1$, but a few had $\Delta$ as high as $10$ or higher. An explicit check of these instances showed that the cause appeared to be freezing spins ($|m_i|\rightarrow 1 $). When removing the few runs having any empirical magnetizations $|m_i|>0.999$, we saw the same modest increase in error as for weak interactions (dashed curve in Figure~\ref{fig:highSampleSimulations}c)). Interestingly, even for runs where the error exploded, most parameters were still correctly identified, \textit{i.e.} a few parameters diverging did not appear to destroy the reconstruction throughout the rest of the system.

In the default set-up we sampled the configurations at time steps spaced $10 \cdot N$ MC steps apart. To assay robustness against such correlations, we also used a spacing $k \cdot N$ and lowered $k$ successively in the strongly correlated case $N=64$, $T=0.5$, $10^6$ samples from Figure~\ref{fig:highSampleSimulations}a) (data not shown). We then saw practically no effect lowering $k$ from $10$ to $1$, while we did see effects at $k\approx 0.5$ and smaller. 

The consistency result (\ref{cons}) assumes an Ising model as the true underlying distribution, a premise which is, at best, only approximately true for any real data set. An interesting question is therefore: will our method deviate markedly from an exact max-entropy inference, as found by Boltzmann learning, when the Ising assumption is not true? The answer seems to be no. To investigate, we imposed an Ising model on a small system (thus feasible for Boltzmann learning) which had both second and third order interactions. Figure~\ref{fig:highSampleSimulations}d) shows that, even when the true system has no pairwise interactions at all, pseudolikelihood maximization and Boltzmann learning estimate very similar "would be" $J_{ij}$:s.
In conclusion, the algorithm appears to be tolerant toward approximate distribution assumptions, high biases as well as data interdependency.


\begin{figure}[t]
\centering
\includegraphics{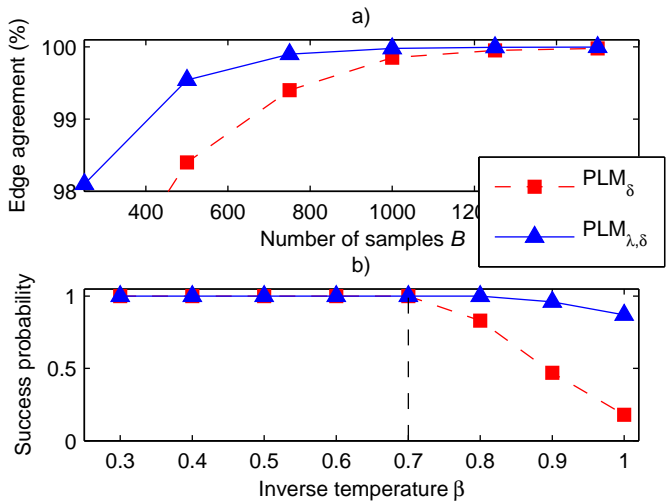}
\caption{(Color online) a) Edge agreement versus sample size in a binary SK model of size $N=100$ and sparsity $p=0.05$ for $PLM_{\delta}$ and $PLM_{\delta,\lambda}$. $T=2$ for all data points. b) Probability of 100\% edge agreement versus inverse temperature for $PLM_{\delta}$ and $PLM_{\delta,\lambda}$ using $B=4500$ on 7x7 nearest-neighbor grids ($N=49$) with 30\% dilution.}
\label{fig:lowSampleSimulations}
\end{figure} 

{\it Results for low-quality data:}
Rebuilding the sign-sparsity pattern of $\boldsymbol{\mathrm{J}}$ from few samples using the pseudolikelihood maximization (PLM) idea has been done numerically for various sparsity types in \cite{RavikumarWainwrightLafferty10} and \cite{BentoMontanari09}. We provide here some additional results, specifically regarding the advantages of using a regularization term. Taking $\lambda>0$ after all makes the optimization problem considerably harder computationally. A simpler approach would be to minimize (\ref{f_def}) with $\lambda=0$ and declare all couplings for which $|J_{ij}|<\delta$ to be zero (for some tolerance $\delta$). Intuitively, inclusion of a regularization term should allow for better utilization of sample information than the simpler tolerance approach. As a test case we look at a version of the SK model where the couplings are not Gaussian but binary, $J_{ij}=\pm \frac{1}{\sqrt{pN}}$ (with equal probability). The inference quality is measured as the percentage of pairs $(i,j)$ where the interaction strength is identified correctly as $"+"$, $"0"$ or $"-"$. PLM using tolerance only and PLM using regularization (as well as a tolerance limit) will be referred to as $PLM_{\delta}$ and $PLM_{\delta,\lambda}$ respectively. Figure \ref{fig:lowSampleSimulations}a) shows that for $N=100$, $p=0.05$ and $T=2$, $PLM_{\delta,\lambda}$ fits the edges more accurately and gives perfect reconstruction for fewer samples than $PLM_{\delta}$. Note that in this example guessing $J^*_{ij}=0$ for all pairs would result in a 95\% edge agreement on average. Optimal values of $\delta$ and $\{\delta,\lambda\}$ for each $B$ were determined empirically and used on 20 new parameter sets to yield the averages.

For several sparsity structures the performance of PLM has been shown to drop as the temperature goes below some $T_{crit}$ even if $B$ is quite large \cite{BentoMontanari09}. One such example is $B=4500$ on 7x7 nearest-neighbor grids with positive couplings, where each edge in the grid is removed with probability 0.3 and the remaining couplings are set to one. The "failure" occurs close to the known critical point for the Ising model on such grids \cite{BentoMontanari09}, $\beta_{crit} \approx 0.7$ \cite{Zobin78}. We applied $PLM_{\delta,\lambda}$ and $PLM_{\delta}$ to this problem to see whether combined regularization-tolerance can boost performance at low temperatures. Figure \ref{fig:lowSampleSimulations}b) shows the outcome, where optimal $\delta$ and $\{\delta,\lambda\}$ for each $\beta$ were again found empirically and probabilities estimated using 200 new grids. A breakdown is indeed seen for $PLM_{\delta}$ around $\beta=0.7$, but the effect on $PLM_{\delta,\lambda}$ is much less pronounced. Perfect edge recovery, using the latter, is had with high probability far into the low-temperature region. The complete data output (not reported) shows that including the tolerance threshold in $PLM_{\delta,\lambda}$ (as opposed to trusting the regularization term alone to force suitable estimates of $J_{ij}$ to zero), becomes necessary at low temperatures. MC samples in this case were generated using a warmup time of $10^7 \cdot N$ spin updates and a sampling frequency of one observation every $2000 \cdot N$ updates.

{\it Discussion:} 
Pseudolikelihood and approximate max-entropy
should be considered alternative approaches, where in both cases exact inference (likelihood or max-entropy) has
been traded in for computability. 
Our results suggest that the pseudolikelihood approach allows for accurate
inference in Ising models even for large strongly coupled systems, a capability which appears to be maintained even when the amount of data is severely limited. Tolerating high external fields, dependence among samples, sparseness as well as strong correlations, in addition to being robust when distribution assumptions are but approximate, the method provides a very complete and real alternative to current approaches that typically fail in one or several of these respects.
Our results also confirm that including an $l_1$ regularization term is helpful in retrieving sign-sparsity from few samples, allowing for complete graph reconstruction even in low-temperature regions.


The PLM objective function has one term per sampled configuration, so running time is heavily dependent on sample size. For instance, the $N=64$ cases with $10^8$ samples took hours on a (one-core) standard home PC using Newton decent, with almost all of the time spent evaluating the Hessian of the objective function (which depends on all $10^8$ samples). We note for future work that one may alternatively use a quasi-Newton or a Conjugate Gradient method, \textit{i.e.} algorithms that don't use (exact) Hessians (initial trials suggest that these work also in the strongly correlated cases). When the number of samples is small, however, the algorithm naturally runs fast. Thus, in the region where pseudolikelihood maximization is likely most interesting (small sample size), it is also computationally efficient and competitive. Moreover, a practical convenience of PLM is that the $N$ subproblems can be solved completely independently, allowing for straightforward parallel execution.

\section*{Acknowledgements} 
E.A. thanks Martin Wainwright, Toshiyuki Tanaka and Michael
H\"ornqvist for useful discussions.
This work was supported by the
Academy of Finland as part of its Finland Distinguished
Professor program, Project No. 129024/Aurell.

\bibliography{spinglas}

\begin{thebibliography}{10}%
\makeatletter
\providecommand \@ifxundefined [1]{%
 \ifx #1\undefined \expandafter \@firstoftwo
 \else \expandafter \@secondoftwo
\fi
}%
\providecommand \@ifnum [1]{%
 \ifnum #1\expandafter \@firstoftwo
 \else \expandafter \@secondoftwo
\fi
}%
\providecommand \enquote [1]{``#1''}%
\providecommand \bibnamefont  [1]{#1}%
\providecommand \bibfnamefont [1]{#1}%
\providecommand \citenamefont [1]{#1}%
\providecommand\href[0]{\@sanitize\@href}%
\providecommand\@href[1]{\endgroup\@@startlink{#1}\endgroup\@@href}%
\providecommand\@@href[1]{#1\@@endlink}%
\providecommand \@sanitize [0]{\begingroup\catcode`\&12\catcode`\#12\relax}%
\@ifxundefined \pdfoutput {\@firstoftwo}{%
 \@ifnum{\z@=\pdfoutput}{\@firstoftwo}{\@secondoftwo}%
}{%
 \providecommand\@@startlink[1]{\leavevmode\special{html:<a href="#1">}}%
 \providecommand\@@endlink[0]{\special{html:</a>}}%
}{%
 \providecommand\@@startlink[1]{%
  \leavevmode
  \pdfstartlink
   attr{/Border[0 0 1 ]/H/I/C[0 1 1]}%
   user{/Subtype/Link/A<</Type/Action/S/URI/URI(#1)>>}%
  \relax
 }%
 \providecommand\@@endlink[0]{\pdfendlink}%
}%
\providecommand \url  [0]{\begingroup\@sanitize \@url }%
\providecommand \@url [1]{\endgroup\@href {#1}{\urlprefix}}%
\providecommand \urlprefix [0]{URL }%
\providecommand \Eprint[0]{\href }%
\@ifxundefined \urlstyle {%
  \providecommand \doi [1]{doi:\discretionary{}{}{}#1}%
}{%
  \providecommand \doi [0]{doi:\discretionary{}{}{}\begingroup
  \urlstyle{rm}\Url }%
}%
\providecommand \doibase [0]{http://dx.doi.org/}%
\providecommand \Doi[1]{\href{\doibase#1}}%
\providecommand \bibAnnote [3]{%
  \BibitemShut{#1}%
  \begin{quotation}\noindent
    \textsc{Key:}\ #2\\\textsc{Annotation:}\ #3%
  \end{quotation}%
}%
\providecommand \bibAnnoteFile [2]{%
  \IfFileExists{#2}{\bibAnnote {#1} {#2} {\input{#2}}}{}%
}%
\providecommand \typeout [0]{\immediate \write \m@ne }%
\providecommand \selectlanguage [0]{\@gobble}%
\providecommand \bibinfo [0]{\@secondoftwo}%
\providecommand \bibfield [0]{\@secondoftwo}%
\providecommand \translation [1]{[#1]}%
\providecommand \BibitemOpen[0]{}%
\providecommand \bibitemStop [0]{}%
\providecommand \bibitemNoStop [0]{.\EOS\space}%
\providecommand \EOS [0]{\spacefactor3000\relax}%
\providecommand \BibitemShut [1]{\csname bibitem#1\endcsname}%
\bibitem{Bialek2006}%
  \BibitemOpen
  \bibfield{author}{%
  \bibinfo {author} {\bibfnamefont{E.}~\bibnamefont{Schneidman}}, \bibinfo
  {author} {\bibfnamefont{M.~J.}\ \bibnamefont{Berry}}, \bibinfo {author}
  {\bibfnamefont{R.}~\bibnamefont{Segev}},\ and\ \bibinfo {author}
  {\bibfnamefont{W.}~\bibnamefont{Bialek}},\ }%
  \bibfield{journal}{%
  \bibinfo {journal} {Nature}\ }%
  \textbf{\bibinfo {volume} {440}},\ \bibinfo {pages} {1007} (\bibinfo {year}
  {2006})%
  \bibAnnoteFile{NoStop}{Bialek2006}%
\bibitem{Lezon2006}%
  \BibitemOpen
  \bibfield{author}{%
  \bibinfo {author} {\bibfnamefont{T.~R.}\ \bibnamefont{Lezon}}, \bibinfo
  {author} {\bibfnamefont{J.~R.}\ \bibnamefont{Banavar}}, \bibinfo {author}
  {\bibfnamefont{M.}~\bibnamefont{Cieplak}}, \bibinfo {author}
  {\bibfnamefont{A.}~\bibnamefont{Maritan}},\ and\ \bibinfo {author}
  {\bibfnamefont{N.~V.}\ \bibnamefont{Fedoroff}},\ }%
  \bibfield{journal}{%
  \bibinfo {journal} {PNAS}\ }%
  \textbf{\bibinfo {volume} {103}},\ \bibinfo {pages} {19033} (\bibinfo {year}
  {2006})%
  \bibAnnoteFile{NoStop}{Lezon2006}%
\bibitem{Cocco2009}%
  \BibitemOpen
  \bibfield{author}{%
  \bibinfo {author} {\bibfnamefont{S.}~\bibnamefont{Cocco}}, \bibinfo {author}
  {\bibfnamefont{S.}~\bibnamefont{Leibler}},\ and\ \bibinfo {author}
  {\bibfnamefont{R.}~\bibnamefont{Monasson}},\ }%
  \bibfield{journal}{%
  \bibinfo {journal} {PNAS}\ }%
  \textbf{\bibinfo {volume} {106}},\ \bibinfo {pages} {14058} (\bibinfo {year}
  {2009})%
  \bibAnnoteFile{NoStop}{Cocco2009}%
\bibitem{Tkacik2009}%
  \BibitemOpen
  \bibfield{author}{%
  \bibinfo {author} {\bibfnamefont{G.}~\bibnamefont{Tkacik}}, \bibinfo {author}
  {\bibfnamefont{E.}~\bibnamefont{Schneidman}}, \bibinfo {author}
  {\bibfnamefont{M.~J.~B.}\ \bibnamefont{II}},\ and\ \bibinfo {author}
  {\bibfnamefont{W.}~\bibnamefont{Bialek}},\ }%
  \bibinfo {note} {arXiv:0912.5409}%
  \bibAnnoteFile{NoStop}{Tkacik2009}%
\bibitem{SchugPNAS2009}%
  \BibitemOpen
  \bibfield{author}{%
  \bibinfo {author} {\bibfnamefont{A.}~\bibnamefont{Schug}}, \bibinfo {author}
  {\bibfnamefont{M.}~\bibnamefont{Weigt}}, \bibinfo {author}
  {\bibfnamefont{J.~N.}\ \bibnamefont{Onuchic}}, \bibinfo {author}
  {\bibfnamefont{T.}~\bibnamefont{Hwa}},\ and\ \bibinfo {author}
  {\bibfnamefont{H.}~\bibnamefont{Szurmant}},\ }%
  \bibfield{journal}{%
  \Doi{10.1073/pnas.0912100106}{\bibinfo {journal} {PNAS}}\ }%
  \textbf{\bibinfo {volume} {106}},\ \bibinfo {pages} {22124} (\bibinfo {year}
  {2009})%
  \bibAnnoteFile{NoStop}{SchugPNAS2009}%
\bibitem{WeigtPNAS2009}%
  \BibitemOpen
  \bibfield{author}{%
  \bibinfo {author} {\bibfnamefont{M.}~\bibnamefont{Weigt}}, \bibinfo {author}
  {\bibfnamefont{R.~A.}\ \bibnamefont{White}}, \bibinfo {author}
  {\bibfnamefont{H.}~\bibnamefont{Szurmant}}, \bibinfo {author}
  {\bibfnamefont{J.~A.}\ \bibnamefont{Hoch}},\ and\ \bibinfo {author}
  {\bibfnamefont{T.}~\bibnamefont{Hwa}},\ }%
  \bibfield{journal}{%
  \Doi{10.1073/pnas.0805923106}{\bibinfo {journal} {PNAS}}\ }%
  \textbf{\bibinfo {volume} {106}},\ \bibinfo {pages} {67} (\bibinfo {year}
  {2009})%
  \bibAnnoteFile{NoStop}{WeigtPNAS2009}%
\bibitem{Morcos-2011}%
  \BibitemOpen
  \bibfield{author}{%
  \bibinfo {author} {\bibfnamefont{F.}~\bibnamefont{Morcos}}, \bibinfo {author}
  {\bibfnamefont{A.}~\bibnamefont{Pagnani}}, \bibinfo {author}
  {\bibfnamefont{B.}~\bibnamefont{Lunt}}, \bibinfo {author}
  {\bibfnamefont{A.}~\bibnamefont{Bertolino}}, \bibinfo {author}
  {\bibfnamefont{D.~S.}\ \bibnamefont{Marks}}, \bibinfo {author}
  {\bibfnamefont{C.}~\bibnamefont{Sander}}, \bibinfo {author}
  {\bibfnamefont{R.}~\bibnamefont{Zecchina}}, \bibinfo {author}
  {\bibfnamefont{J.~N.}\ \bibnamefont{Onuchic}}, \bibinfo {author}
  {\bibfnamefont{T.}~\bibnamefont{Hwa}},\ and\ \bibinfo {author}
  {\bibfnamefont{M.}~\bibnamefont{Weigt}},\ }%
  \bibfield{journal}{%
  \bibinfo {journal} {PNAS}\ }%
  \textbf{\bibinfo {volume} {108}} (\bibinfo {year} {2011})%
  \bibAnnoteFile{NoStop}{Morcos-2011}%
\bibitem{RavikumarWainwrightLafferty10}%
  \BibitemOpen
  \bibfield{author}{%
  \bibinfo {author} {\bibfnamefont{P.}~\bibnamefont{Ravikumar}}, \bibinfo
  {author} {\bibfnamefont{M.~J.}\ \bibnamefont{Wainwright}},\ and\ \bibinfo
  {author} {\bibfnamefont{J.~D.}\ \bibnamefont{Lafferty}},\ }%
  \bibfield{journal}{%
  \bibinfo {journal} {Annals of Statistics}\ }%
  \textbf{\bibinfo {volume} {38}},\ \bibinfo {pages} {1287} (\bibinfo {year}
  {2010})%
  \bibAnnoteFile{NoStop}{RavikumarWainwrightLafferty10}%
\bibitem{darmois35}%
  \BibitemOpen
  \bibfield{author}{%
  \bibinfo {author} {\bibfnamefont{G.}~\bibnamefont{Darmois}},\ }%
  \bibfield{journal}{%
  \bibinfo {journal} {C.R. Acad. Sci. Paris}\ }%
  \textbf{\bibinfo {volume} {200}},\ \bibinfo {pages} {1265} (\bibinfo {year}
  {1935})%
  \bibAnnoteFile{NoStop}{darmois35}%
\bibitem{pitman36}%
  \BibitemOpen
  \bibfield{author}{%
  \bibinfo {author} {\bibfnamefont{E.}~\bibnamefont{Pitman}}\ and\ \bibinfo
  {author} {\bibfnamefont{J.}~\bibnamefont{Wishart}},\ }%
  \bibfield{journal}{%
  \Doi{10.1017/S0305004100019307}{\bibinfo {journal} {Math. Proc. Cambridge}}\
  }%
  \textbf{\bibinfo {volume} {32}},\ \bibinfo {pages} {567} (\bibinfo {year}
  {1936})%
  \bibAnnoteFile{NoStop}{pitman36}%
\bibitem{koopman36}%
  \BibitemOpen
  \bibfield{author}{%
  \bibinfo {author} {\bibfnamefont{B.}~\bibnamefont{Koopman}},\ }%
  \bibfield{journal}{%
  \Doi{10.2307/1989758}{\bibinfo {journal} {T. Am. Math. Soc.}}\ }%
  \textbf{\bibinfo {volume} {39}},\ \bibinfo {pages} {399} (\bibinfo {year}
  {1936})%
  \bibAnnoteFile{NoStop}{koopman36}%
\bibitem{Ackley85alearning}%
  \BibitemOpen
  \bibfield{author}{%
  \bibinfo {author} {\bibfnamefont{H.}~\bibnamefont{Ackley}}, \bibinfo {author}
  {\bibfnamefont{E.}~\bibnamefont{Hinton}},\ and\ \bibinfo {author}
  {\bibfnamefont{J.}~\bibnamefont{Sejnowski}},\ }%
  \bibfield{journal}{%
  \bibinfo {journal} {Cogn. Sci.}\ }%
  \textbf{\bibinfo {volume} {9}},\ \bibinfo {pages} {147} (\bibinfo {year}
  {1985})%
  \bibAnnoteFile{NoStop}{Ackley85alearning}%
\bibitem{Broderick}%
  \BibitemOpen
  \bibfield{author}{%
  \bibinfo {author} {\bibfnamefont{T.}~\bibnamefont{Broderick}}, \bibinfo
  {author} {\bibfnamefont{M.}~\bibnamefont{Dudik}}, \bibinfo {author}
  {\bibfnamefont{G.}~\bibnamefont{Tkacik}}, \bibinfo {author}
  {\bibfnamefont{R.~E.}\ \bibnamefont{Schapire}},\ and\ \bibinfo {author}
  {\bibfnamefont{W.}~\bibnamefont{Bialek}}}%
   (\bibinfo {year} {2007}),\ \bibinfo {note} {arXiv:0712.2437}%
  \bibAnnoteFile{NoStop}{Broderick}%
\bibitem{CarreiraHinton}%
  \BibitemOpen
  \bibfield{author}{%
  \bibinfo {author} {\bibfnamefont{M.~A.}\ \bibnamefont{Carreira-Perpinan}}\
  and\ \bibinfo {author} {\bibfnamefont{G.~E.}\ \bibnamefont{Hinton}},\ }%
  \bibinfo {note} {from Artificial Intelligence and Statistics, 2005,
  Barbados}%
  \bibAnnoteFile{NoStop}{CarreiraHinton}%
\bibitem{Frontiers}%
  \BibitemOpen
  \bibfield{author}{%
  \bibinfo {author} {\bibfnamefont{Y.}~\bibnamefont{Roudi}}, \bibinfo {author}
  {\bibfnamefont{J.~A.}\ \bibnamefont{Hertz}},\ and\ \bibinfo {author}
  {\bibfnamefont{E.}~\bibnamefont{Aurell}},\ }%
  \bibfield{journal}{%
  \bibinfo {journal} {Front. Comput. Neurosci.}\ }%
  \textbf{\bibinfo {volume} {3}} (\bibinfo {year} {2009})%
  \bibAnnoteFile{NoStop}{Frontiers}%
\bibitem{Rodriguez97efficientlearning}%
  \BibitemOpen
  \bibfield{author}{%
  \bibinfo {author} {\bibfnamefont{H.}~\bibnamefont{Kappen}}\ and\ \bibinfo
  {author} {\bibfnamefont{F.~B.}\ \bibnamefont{Rodriguez}},\ }%
  \bibfield{journal}{%
  \bibinfo {journal} {Neural Comput.}\ }%
  \textbf{\bibinfo {volume} {10}},\ \bibinfo {pages} {1137} (\bibinfo {year}
  {1998})%
  \bibAnnoteFile{NoStop}{Rodriguez97efficientlearning}%
\bibitem{SessakMonasson}%
  \BibitemOpen
  \bibfield{author}{%
  \bibinfo {author} {\bibfnamefont{V.}~\bibnamefont{Sessak}}\ and\ \bibinfo
  {author} {\bibfnamefont{R.}~\bibnamefont{Monasson}},\ }%
  \bibfield{journal}{%
  \bibinfo {journal} {J. Phys. A: Math. Theor.}\ }%
  \textbf{\bibinfo {volume} {42}} (\bibinfo {year} {2009})%
  \bibAnnoteFile{NoStop}{SessakMonasson}%
\bibitem{MezardMora}%
  \BibitemOpen
  \bibfield{author}{%
  \bibinfo {author} {\bibfnamefont{M.}~\bibnamefont{M\'ezard}}\ and\ \bibinfo
  {author} {\bibfnamefont{T.}~\bibnamefont{Mora}},\ }%
  \bibfield{journal}{%
  \bibinfo {journal} {J. Physiology-Paris}\ }%
  \textbf{\bibinfo {volume} {103}},\ \bibinfo {pages} {107} (\bibinfo {year}
  {2009})%
  \bibAnnoteFile{NoStop}{MezardMora}%
\bibitem{Marinari}%
  \BibitemOpen
  \bibfield{author}{%
  \bibinfo {author} {\bibfnamefont{E.}~\bibnamefont{Marinari}}\ and\ \bibinfo
  {author} {\bibfnamefont{V.~V.}\ \bibnamefont{Kerrebroeck}},\ }%
  \bibfield{journal}{%
  \bibinfo {journal} {J. Stat. Mech.}}%
   (\bibinfo {year} {2010})%
  \bibAnnoteFile{NoStop}{Marinari}%
\bibitem{AurellOllionRoudi10}%
  \BibitemOpen
  \bibfield{author}{%
  \bibinfo {author} {\bibfnamefont{E.}~\bibnamefont{Aurell}}, \bibinfo {author}
  {\bibfnamefont{C.}~\bibnamefont{Ollion}},\ and\ \bibinfo {author}
  {\bibfnamefont{Y.}~\bibnamefont{Roudi}},\ }%
  \bibfield{journal}{%
  \bibinfo {journal} {Eur. Phys. J. B}\ }%
  \textbf{\bibinfo {volume} {77}} (\bibinfo {year} {2010})%
  \bibAnnoteFile{NoStop}{AurellOllionRoudi10}%
\bibitem{CoccoMonasson2011}%
  \BibitemOpen
  \bibfield{author}{%
  \bibinfo {author} {\bibfnamefont{S.}~\bibnamefont{Cocco}}\ and\ \bibinfo
  {author} {\bibfnamefont{R.}~\bibnamefont{Monasson}},\ }%
  \bibfield{journal}{%
  \bibinfo {journal} {Phys. Rev. Lett.}\ }%
  \textbf{\bibinfo {volume} {106}},\ \bibinfo {pages} {090601} (\bibinfo
  {month} {Mar}\ \bibinfo {year} {2011})%
  \bibAnnoteFile{NoStop}{CoccoMonasson2011}%
\bibitem{PhysRevLett.107.220601}%
  \BibitemOpen
  \bibfield{author}{%
  \bibinfo {author} {\bibfnamefont{J.}~\bibnamefont{Sohl-Dickstein}}, \bibinfo
  {author} {\bibfnamefont{P.~B.}\ \bibnamefont{Battaglino}},\ and\ \bibinfo
  {author} {\bibfnamefont{M.~R.}\ \bibnamefont{DeWeese}},\ }%
  \bibfield{journal}{%
  \bibinfo {journal} {Phys. Rev. Lett.}\ }%
  \textbf{\bibinfo {volume} {107}},\ \bibinfo {pages} {220601} (\bibinfo {year}
  {2011})%
  \bibAnnoteFile{NoStop}{PhysRevLett.107.220601}%
\bibitem{Tibshirani96}%
  \BibitemOpen
  \bibfield{author}{%
  \bibinfo {author} {\bibfnamefont{R.}~\bibnamefont{Tibshirani}},\ }%
  \bibfield{journal}{%
  \bibinfo {journal} {J. Roy. Stat. Soc., Ser. B}\ }%
  \textbf{\bibinfo {volume} {58}},\ \bibinfo {pages} {267} (\bibinfo {year}
  {1996})%
  \bibAnnoteFile{NoStop}{Tibshirani96}%
\bibitem{DonohoElad03}%
  \BibitemOpen
  \bibfield{author}{%
  \bibinfo {author} {\bibfnamefont{D.}~\bibnamefont{Donoho}}\ and\ \bibinfo
  {author} {\bibfnamefont{M.}~\bibnamefont{Elad}},\ }%
  \bibfield{journal}{%
  \bibinfo {journal} {Proc. Natl. Acad. Sci. USA}\ }%
  \textbf{\bibinfo {volume} {100}},\ \bibinfo {pages} {2197} (\bibinfo {year}
  {2003})%
  \bibAnnoteFile{NoStop}{DonohoElad03}%
\bibitem{NYAS:NYAS03764}%
  \BibitemOpen
  \bibfield{author}{%
  \bibinfo {author} {\bibfnamefont{M.}~\bibnamefont{Gustafsson}}, \bibinfo
  {author} {\bibfnamefont{M.}~\bibnamefont{H\"{o}rnquist}}, \bibinfo {author}
  {\bibfnamefont{J.}~\bibnamefont{Lundstr\"{o}m}}, \bibinfo {author}
  {\bibfnamefont{J.}~\bibnamefont{Bj\"{o}rkegren}},\ and\ \bibinfo {author}
  {\bibfnamefont{J.}~\bibnamefont{Tegn\'er}},\ }%
  \bibfield{journal}{%
  \bibinfo {journal} {Ann. N.Y. Acad. Sci.}\ }%
  \textbf{\bibinfo {volume} {1158}},\ \bibinfo {pages} {265} (\bibinfo {year}
  {2009})%
  \bibAnnoteFile{NoStop}{NYAS:NYAS03764}%
\bibitem{KohKimBoyd07}%
  \BibitemOpen
  \bibfield{author}{%
  \bibinfo {author} {\bibfnamefont{K.}~\bibnamefont{Koh}}, \bibinfo {author}
  {\bibfnamefont{S.}~\bibnamefont{Kim}},\ and\ \bibinfo {author}
  {\bibfnamefont{S.}~\bibnamefont{Boyd}},\ }%
  \bibfield{journal}{%
  \bibinfo {journal} {J. Mach. Learn. Res.}\ }%
  \textbf{\bibinfo {volume} {3}},\ \bibinfo {pages} {1519} (\bibinfo {year}
  {2007})%
  \bibAnnoteFile{NoStop}{KohKimBoyd07}%
\bibitem{SherringtonKirkpatrick75}%
  \BibitemOpen
  \bibfield{author}{%
  \bibinfo {author} {\bibfnamefont{D.}~\bibnamefont{Sherrington}}\ and\
  \bibinfo {author} {\bibfnamefont{S.}~\bibnamefont{Kirkpatrick}},\ }%
  \bibfield{journal}{%
  \bibinfo {journal} {Phys. Rev. Lett}\ }%
  \textbf{\bibinfo {volume} {35}},\ \bibinfo {pages} {1792} (\bibinfo {year}
  {1975})%
  \bibAnnoteFile{NoStop}{SherringtonKirkpatrick75}%
\bibitem{BentoMontanari09}%
  \BibitemOpen
  \bibfield{author}{%
  \bibinfo {author} {\bibfnamefont{J.}~\bibnamefont{Bento}}\ and\ \bibinfo
  {author} {\bibfnamefont{A.}~\bibnamefont{Montanari}},\ }%
  \bibfield{journal}{%
  \bibinfo {journal} {NIPS}\ }%
  \textbf{\bibinfo {volume} {22}} (\bibinfo {year} {2009})%
  \bibAnnoteFile{NoStop}{BentoMontanari09}%
\bibitem{Zobin78}%
  \BibitemOpen
  \bibfield{author}{%
  \bibinfo {author} {\bibfnamefont{D.}~\bibnamefont{Zobin}},\ }%
  \bibfield{journal}{%
  \bibinfo {journal} {Phys. Rev. B}\ }%
  \textbf{\bibinfo {volume} {18}},\ \bibinfo {pages} {2387} (\bibinfo {year}
  {1978})%
  \bibAnnoteFile{NoStop}{Zobin78}%
\end{thebibliography}%
\end{document}